\documentclass[12pt]{iopart}

\usepackage{iopams}  
\usepackage{graphicx}
\begin{document}

\title[]{Storage and retrieval of nonclassical photon pairs and conditional single photons generated by parametric down-conversion process}

\author{K Akiba$^1$
, K Kashiwagi$^1$
, M Arikawa$^1$
, M Kozuma$^{1,2}$} 
\address{%
$^1$Department of Physics, Tokyo Institute of Technology, 
2-12-1 O-okayama, Meguro-ku, Tokyo 152-8550, Japan \\
$^2$CREST, Japan Science and Technology Agency, 1-9-9 Yaesu, Chuo-ku, Tokyo 103-0028, Japan
}
\begin{abstract}
Storage and retrieval of parametric down-conversion (PDC) photons are demonstrated with electromagnetically induced transparency (EIT). 
Extreme frequency filtering is performed for THz order of broadband PDC light and the frequency bandwidth of the light is reduced to MHz order. 
Storage and retrieval procedures are carried out for the frequency filtered PDC photons. Since the filtered bandwidth [full width at half-maximum (FWHM) = 9 MHz] is within the EIT window (FWHM = 12.6 MHz), the flux of the PDC light is successfully stored and retrieved. 
The nonclassicality of the retrieved light is confirmed by using photon counting method, where the classical inequality which is only satisfied for classical light fields is introduced. 
Since the PDC photons can be utilized for producing the single photon state conditionally, storage and retrieval procedures are also performed for the conditional single photons. 
Anti-correlation parameter used for checking the property of single photon state shows the value less than 1, which means the retrieved light is in a non-classical region. 

\end{abstract}


\maketitle

\section{Introduction}
Coherently transferring a quantum state of light to an atomic ensemble enables us to implement quantum memory for photons \cite{DSP,QM}, generate a nonclassical state of an atomic ensemble, and manipulate a quantum state of light \cite{SoL2,manipulate1,manipulate2}. A coherent transfer can be made by using a phenomenon called electromagnetically induced transparency (EIT) \cite{EIT1}. Recently, storage and retrieval of nonclassical light and generation of entanglement between two distant atomic ensembles have been demonstrated \cite{Kuzmich, Lukin, remote}. In these experiments, Raman scattered photons generated from atomic ensembles were used as nonclassical light \cite{nonclassical}. Another method of generating nonclassical light, the parametric down-conversion (PDC) process, has been widely used in fundamental experiments in quantum optics \cite{Mandel} and for demonstrating various quantum information protocols \cite{QT,QCR}. Light fields produced by the PDC process have been used as rich nonclassical photonic states such as conditional single photon state, antibunched state \cite{Koashi,Ou}, 
hyperentangled state \cite{Hyperentangle}
GHZ state \cite{GHZ}, 
$W$ state \cite{W},
and cluster state \cite{cluster}, that is, the PDC photon pair is a fundamental resource for creating various nonclassical light fields. Storing the PDC photons in atomic ensemble, therefore, enables us to store and manipulate various nonclassical lights and can be utilized in the present photonic quantum information processing. 

Storing nonclassical light generated from the PDC process, however, has a problem in that the bandwidth of the light is not matched by the atomic interaction width \cite{A}. The spectral bandwidth of the PDC photons is typically the THz order, whereas the spectral widths for the atomic dipolar transition is the MHz order. This problem becomes crucial for photon counting, since a photon interacting with atoms cannot be selectively counted. Frequency filtering of PDC photons is thus required to solve this problem \cite{B}. It should be noted that frequency filtering does not impose fundamental limitations to the current protocols on photonic quantum information, while it does have a disadvantage in that the filtering decreases the photon counting rate. The storage of PDC photons has been demonstrated by using time-filtering in our previous study \cite{FSPF}. However, the nonclassicality of the incident and the retrieved lights was not confirmed in the study. Furthermore, there was a large difference of frequency bandwidth between the incident and the retrieved lights.

In this paper, we demonstrated storage and retrieval of PDC photons, where we avoided the problem on the bandwidth mismatch between PDC photons and EIT spectral width by using frequency filtering technique. Preservation of nonclassicality of the retrieved PDC photons was confirmed by using the photon counting method, where we introduced an inequality which is only satisfied for classical light fields. The PDC photons exhibit pair correlation, since a single pump photon is converted to two photons. This pair correlation is essential for preparing single photons conditionally. Therefore, we also checked preservation of the single photon property through the storage process by using triggered PDC photons.

\section{Method}
In our experiment, we used single-mode PDC photons (degenerate parametric fluorescence) because of the experimental simplicity of the frequency filtering. There are several methods for estimating the nonclassicality of the single-mode PDC photons \cite{nonclassical1, nonclassical2, nonclassical3, Waks}. Especially in Ref \cite{Waks}, nonclassical photon statistics has been successfully observed by using photon counting. This method, however, cannot be applied to our experiment because of the low overall photon detection efficiency which is caused by frequency filtering, EIT, and photon counters. We thus introduce a different type of inequality here. When the Glauber's P function \cite{Cauchy-Schwarz} $P(\alpha)$ is non-negative, the following inequality is obtained. 

\begin{equation}
\int \frac{(|\alpha|^4 t -|\alpha|^2)^2}{|\alpha|^2}  P(\alpha) d^2\alpha \ge 0, 
\end{equation} where $t$ is any real number. Left part can be expanded in series on $t$. 
To satisfy the inequality for any $t$, 
the discriminant should be negative and the following inequality is obtained. 

\begin{equation}
w \equiv \frac{\langle\hat{a}^\dagger\hat{a}
\rangle\langle \hat{a}^\dagger\hat{a}^\dagger\hat{a}^\dagger\hat{a}\hat{a}\hat{a}\rangle}
{\langle\hat{a}^\dagger\hat{a}^\dagger\hat{a}\hat{a}\rangle^2} \ge 1, \label{w}
\end{equation} where $\hat{a}$ and $\hat{a}^\dagger$ are the annihilation and creation operators of the single-mode light. 
The violation of this inequality (2) means that the non-negative P function does not exist, that is, the light is nonclassical.

In our experiment, we also prepared single photon state by using the single-mode PDC photons. 
The single photon property can be estimated from the anticorrelation parameter \cite{anti1,anti2}. 

\begin{equation}
\alpha =\frac{\langle\hat{a}_t^\dagger\hat{a}_t
\rangle\langle \hat{a}_t^\dagger\hat{a}_t\hat{a}_s^\dagger\hat{a}_s^\dagger\hat{a}_s\hat{a}_s\rangle}
{\langle\hat{a}_t^\dagger\hat{a}_t\hat{a}_s^\dagger\hat{a}_s\rangle^2}, \label{anti}
\end{equation} where $\hat{a}_s (\hat{a}_s^\dagger)$ and $\hat{a}_t (\hat{a}_t^\dagger)$ are the annihilation (creation) operators of the signal and trigger photons, respectively. The anticorrelation parameter $\alpha$ corresponds to conditionally estimated intensity correlation function. In this sense, when $\alpha$ is less than 1, the resultant signal light is a nonclassical field. Especially, when $\alpha$ is close to 0, the resultant signal light can be regarded as the conditional single photon state. 
When the single-mode parametric fluorescence with low excitation is split into two beams (trigger and signal beams) and one photon is detected at the trigger beam, the resultant signal light can also be regarded as the conditional single photon state. 

\section{Experiment}
\subsection{Storage and retrieval of the frequency-filtered PDC photons}
\begin{figure}
 \begin{center}
 \includegraphics[width=0.6\linewidth]{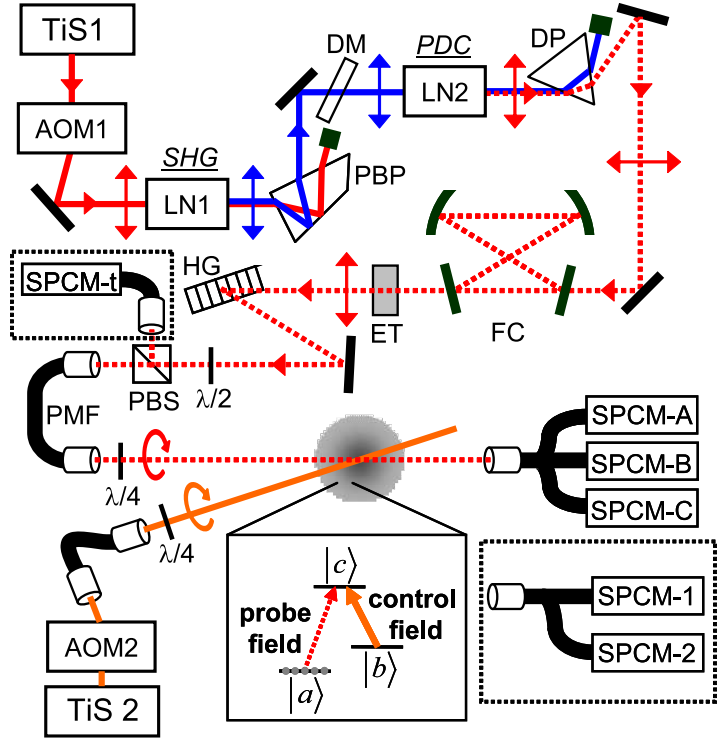}
  \caption{\label{fig:expsetup}
   Schematic diagram of experimental setup. TiS: Ti:sapphire laser; AOM: acousto-optic modulator; LN: quasi-phase-matched MgO:LiNbO$_3$ wave guide; PBP: Pellin--Broca prism; DM: dichroic mirror; DP: dispersing prism; FC: filtering cavity; ET: etalon; HG: holographic grating; $\lambda/2$, $\lambda/4$: half and quarter wave plates; PBS: polarizing beam splitter; PMF: polarization maintaining single-mode fiber; SPCM: single photon counting module. The arrows indicate the polarization of light.}
 \end{center}
\end{figure}

Fig. \ref{fig:expsetup} shows a schematic diagram of the experimental setup. 
The states \{$|{a}\rangle$, $|{b}\rangle$\} and $|{c}\rangle$ 
of a $\Lambda$-type three-level system correspond to $5^2S_{1/2},\ F=\{1, 2\}$ 
and $5^2P_{1/2},\ F'=2$ of the $^{87}$Rb $D_1$ line, respectively. 
We generated PDC photons by using two independent type-0 quasi-phase matched MgO:LiNbO$_3$ waveguides (LN1,2). We employed ridge waveguides 5.0$\mu$m wide $\times$ 3.0 $\mu$m thick $\times$ 8.5mm long, which were fabricated so that quasi-phase-matching was obtained at room temperature. Optical pulse (full width at half-maximum (FWHM) of the pulse temporal width = 50 ns) were generated by an acousto-optic modulator (AOM1) from the linearly polarized continuous output of a Ti:sapphire laser (TiS1) tuned to the $|{a}\rangle\to|{c}\rangle$ transition. The pulses were coupled to LN1 with 60\% efficiency and generated second harmonic pulses, the polarizations of which were the same as the fundamental light. The second harmonic light were separated from the fundamental light by using a Pellin Broca prism (PBP) and a dichroic mirror (DM). Only the second harmonic light was injected into LN2 (coupling efficiency 45 \%) and the PDC light was generated thought the spontaneous PDC process, where the polarization of the PDC light was the same as the second harmonic light. The PDC light was separated from the second harmonic light by using a dispersing prism (DP).
We reduced the bandwidth of the PDC photons from 10 THz to 9 MHz 
by using a filtering cavity (FC) [FWHM of the bandwidth = 9 MHz, free spectral range (FSR) = 1.5 GHz], 
an etalon (ET) [FWHM of the bandwidth = 300 MHz, FSR = 37 GHz], and 
a combination of a holographic grating (HG) and a polarization maintaining single mode fiber (PMF) [FWHM of the bandwidth = 23 GHz]. The total transmittance of the frequency filtering at the center frequency of the PDC light (resonant on $|{a}\rangle\to|{c}\rangle$ transition) was about 20\%.

In order to achieve the long term stability of the frequency filtering, we utilized both the locking and the alignment beams (not shown in Fig.1). A part of the output beam from TiS1 (locking beam) was phase modulated by an electro-optic modulator and its frequency was adequately shifted by an acousto-optic modulator. The locking beam was incident on FC and the resonant frequency of FC was actively stabilized to $| a\rangle \to | c\rangle$ transition by using a piezo electric transducer attached to the cavity mirror, where conventional FM-sideband locking technique was utilized and the fluctuation of the resonant frequency was suppressed less than 1 MHz. In order to prevent the locking beam being mixed to the PDC light, the beams were in opposite circulations of the cavity. 
A fraction of the laser beam diffracted by the AOM1 was tapped by a beam splitter and was used as the alignment beam. The alignment beam was coupled to LN2 so that its spatial mode was made identical with the PDC light. The resonant frequency of the ET was tuned to  $| a\rangle \to | c\rangle$  transition by monitoring the transmittance of the alignment beam and adjusting the temperature of the ET.  Once the proper resonance was obtained, it was stable for one day and the alignment beam was thus cut off when the data was acquired for the PDC light. When it was needed to take data for coherent state of light, the alignment beam was utilized instead of the PDC light.

Magneto-optically trapped $^{87}$Rb atoms were used as an optically thick medium, 
placed in a vacuum cell magnetically shielded by a single permalloy. 
One cycle of the experiment consisted of 
an atomic medium preparation period (1.5 ms) and a measurement period (1.0 ms). 
After laser cooling, the magnetic field and the cooling (-20MHz detuned from $5^2S_{1/2},\ F=2 \to 5^2P_{3/2},\ F'=3$) and repumping ($5^2S_{1/2},\ F=1 \to 5^2P_{3/2},\ F'=2$) lights were turned off, 
and depumping lights ($5^2S_{1/2},\ F=2 \to 5^2P_{3/2},\ F'=2$) illuminated the atomic ensemble for 100 $\mu$s so that all the atoms were prepared in state $|{a}\rangle$, where the optical depth of the $|{a}\rangle\to|{c}\rangle$ transition was $\sim$7. 
In the measurement period, a probe light (beam diameter 250 $\mu$m) and a 250 $\mu$W control light (beam diameter 500 $\mu$m) with the same circular polarizations were injected into the unpolarized atomic ensemble. 
The probe and the control lights were overlapped at the cold atoms with a crossing angle of 3$^\circ$. 
The control light was generated from the TiS2 and 
its intensity was varied using an acousto-optic modulater (AOM2). 
The probe light was detected by single-photon counting modules using silicon avalanche photodiodes 
(SPCM, Perkin--Elmer model SPCM-AQR, detection efficiency: 62\%). 

\begin{figure}
 \begin{center}
 \includegraphics[width=0.5\linewidth]{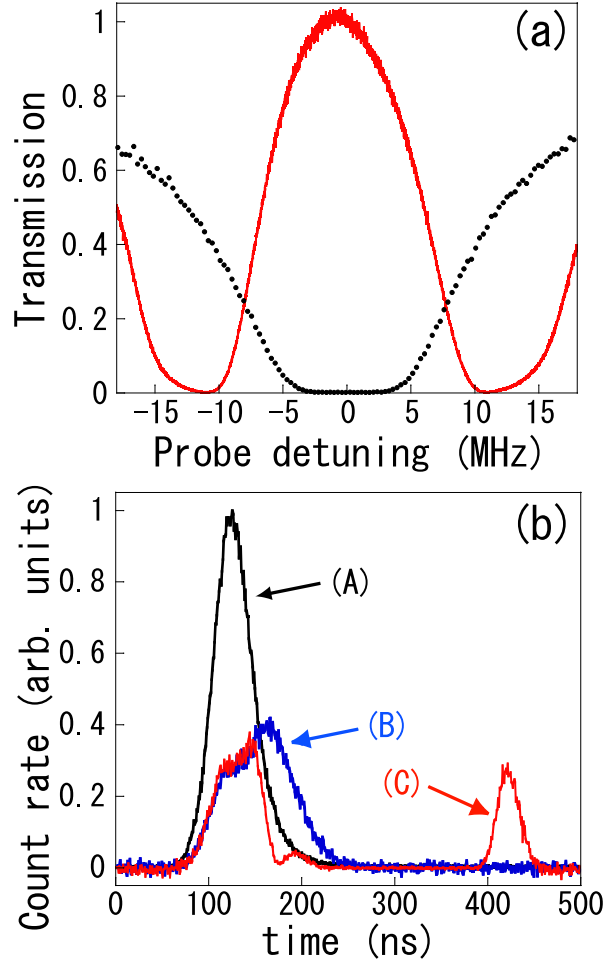}
  \caption{\label{fig:EIT} (a) Measured intensity transmission spectra 
of a weak coherent probe as a function of probe detuning to the $|a\rangle \to |c\rangle$ transition with (solid line) and 
without (dotted line) the control light. 
(b) Typical storage and retrieval of PDC photons: (A) incident, (B) slowed and (C) stored and retrieved pulses.}
 \end{center} 
\end{figure}
Fig. \ref{fig:EIT}(a) shows intensity transmission spectra 
when a weak ($\sim$pW) laser light (alignment beam; coherent state of light) was used as the probe light.
In the absence of the control light (dotted line), the probe laser light was absorbed by the atomic medium.
With the addition of the control light (solid line), 
the atomic medium was rendered transparent around the resonant frequency 
(peak transmission: about 100\%, FWHM of the transparency window: 12.6 MHz ). 
Fig. \ref{fig:EIT}(b) shows typical experimental results of 
the storage and retrieval of frequency-filtered PDC photons. The probe PDC pulses were injected into the cold atoms 1817 times during the measurement period.
Repeating the preparation and measurement periods, we accumulated photon counts, shown as a function of time. Curve (A) corresponds to the probe PDC pulse in the absence of cold atoms.
The temporal shape of the probe PDC pulse was Gaussian with an FWHM of 50 ns. 
Curve (B) corresponds to 
a probe PDC pulse with cold atoms and a constant intensity of the control light. Here, the offset caused by the control light (The unwanted photon flux was mixed to the retrieved light, which was the leakage of the control light and also the spontaneous emission from the background gas.) was subtracted. 
It should be noted that the spectrum of the frequency-filtered PDC photons is not Gaussian but close to Lorentzian, since the probe PDC pulse passed through the filtering cavity (FC). A Lorentzian function has a long tail and therefore the probe PDC pulse included a frequency component which did not interact with atoms. This component passed through the cold atoms at the speed of light in vacuum. In contrast, the frequency component within the EIT window exhibited a slow propagation. 
The observed pulse delay of the component is about 35 ns, corresponding to more than three orders of magnitude reduction in group velocity. 
Due to the contribution of these two frequency components, the temporal shape of the probe PDC pulse was broadened. 
Curve (C) corresponds to the temporal variation of the probe PDC pulse with atoms and 
dynamically changing control light, where 
the control light was turned off at 140 ns and turned on at 390 ns. 
The falling/rising time was 30 ns.  
The total photon count for the retrieved light was 14\% of that for the incident light, which 
clearly shows that the photon flux of the PDC photons was stored and retrieved.

\begin{figure}
 \begin{center}
 \includegraphics[width=0.5\linewidth]{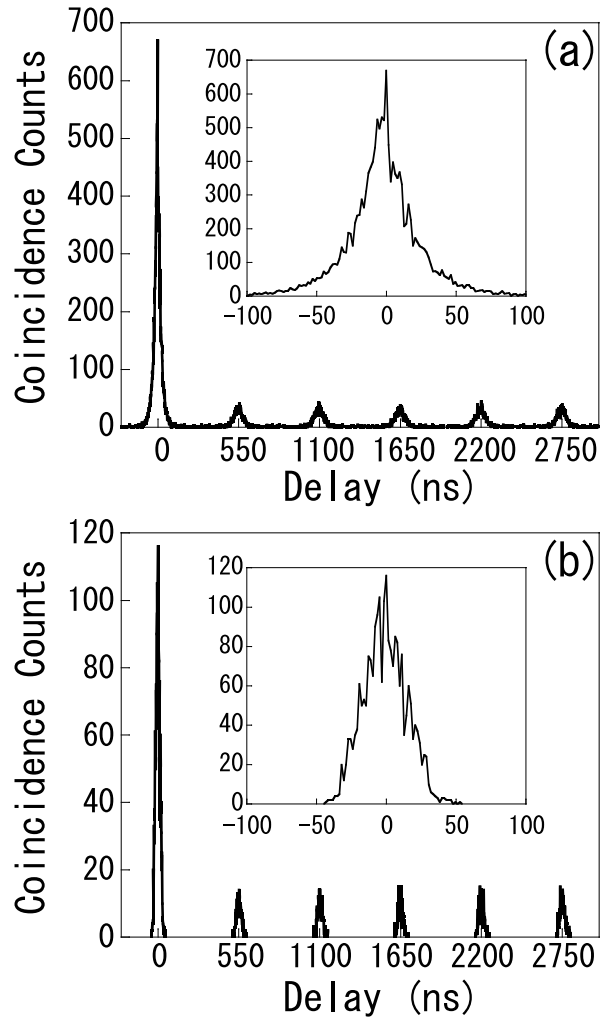}
  \caption{\label{fig:g2} Experimental results of intensity correlation with 1.6 ns resolution of the MCS: (a) incident frequency-filtered PDC photons for 1 hour and (b) retrieved component of the PDC photons for 2 hours. The insets show magnifications of the coincidences around 0 time delay.}
 \end{center}
\end{figure}

\subsection{Estimation of nonclassicality for the PDC photons}
First, we checked the nonclassical property of the retrieved PDC photons, where the rotating angle of the half-wave plate was adjusted so that 
all the probe light passed through the polarizing beam splitter (PBS). The transmitted probe was detected by SPCM-A, B, and C (fig. \ref{fig:expsetup}). 
The total transmissions for frequency-filtered PDC photons at SPCM-A, B, and C were 36\%, 16\%, and 14\%, respectively.
Fig. \ref{fig:g2}(a) shows the auto intensity correlation of incident-frequency-filtered PDC photons 
(curve (A) in fig. \ref{fig:EIT}(b)) obtained by a multi-channel scaler (MCS). 
The obtained normalized autocorrelation function was $g^{(2)}(0)=7.9\pm0.1$, which indicates a superbunching effect, a significant property of PDC photons.
The inset of fig. \ref{fig:g2}(a) shows a magnification of the coincidence counts around 0 time delay.
The FWHM of the correlation time is about 25 ns, 
which agrees with the value obtained from the bandwidth (9 MHz) of the filtering cavity \cite{filterOu}.  
The obtained single counts of SPCM-A was $N_{A}=2968909$. The coincidence counts between SPCM-A and B and between SPCM-A and C were $N_{AB}=18956$ and $N_{AC}=15274$, respectively. The triple counts between SPCM-A, B, and C were 
$N_{ABC}=75$. These were measured for 1700 s. 
The value of $w=N_{A}N_{ABC}/N_{AB}N_{AC}$ was $0.77\pm0.09$.
(The value of $w$ for the incident alignment beam was $1.07\pm0.06$.)
The violation of the inequality (\ref{w}) means that the incident PDC photons 
were in the nonclassical highly bunched photon pair state \cite{FSPF}.

To estimate the retrieved component (curve (C) in fig. \ref{fig:EIT}(b)), 
the outputs of SPCM-A, B, and C were gated from 400 ns to 450 ns. 
While the offset caused by the control beam was subtracted in fig. \ref{fig:EIT}(b), 
such a subtraction cannot be performed for the photon counts. 
The ratio of the retrieved component to the offset in gated time was 2.0.
Fig. \ref{fig:g2}(b) shows the auto intensity correlation of the retrieved light.
The obtained normalized autocorrelation function was $g^{(2)}(0)=7.7\pm0.2$ and 
the superbunching effect was thus preserved \cite{FSPF}.
The value of $w$ for the retrieved light was estimated to be $0.52\pm0.26$ from the obtained values of 
$N_{A}=10420108$,
$N_{AB}=10255$,
$N_{AC}=7782$, and
$N_{ABC}=4$,
which were measured for 10 hours. (The value of $w$ for the retrieved alignment beam was $1.11\pm0.13$.)
The violation of the inequality (\ref{w}) is direct evidence that the nonclassical highly bunched photon pair state was stored and retrieved
(the probability that the retrieved light was in the classical region is 0.05). 
Despite the fact that the retrieved light included the offset component caused by the control light, the values of $g^{(2)}(0)$ and $w$ for the retrieved light were not degraded compared to those for the incident light. The long tail of the Lorentzian spectrum outside the EIT window was not stored, which would have improved the values of $g^{(2)}(0)$ and $w$. On the other hand, the retrieved light had an associated offset which should have degraded these values. We guess that these two effects balanced and hence the values of $g^{(2)}(0)$ and $w$ did not show significant difference between the incident and retrieved lights.

\subsection{Estimation of single photon property for the PDC photons}
Second, we estimated the property of conditional single photons prepared from the PDC photons.
The detection scheme shown in the inset of fig. \ref{fig:expsetup} was used.
A part of the frequency-filtered PDC light was reflected by the PBS, coupled to a single-mode fiber and detected by SPCM-t, where the coupling efficiency to the fiber was 67\%. 
Total transmissions after the PBS were 23\% and 22\% at SPCM-1 and 2, respectively.
The splitting ratio of the PBS was adjusted so that the counting rate of each SPCM became comparable. 

\begin{figure}
 \begin{center}
 \includegraphics[width=0.5\linewidth]{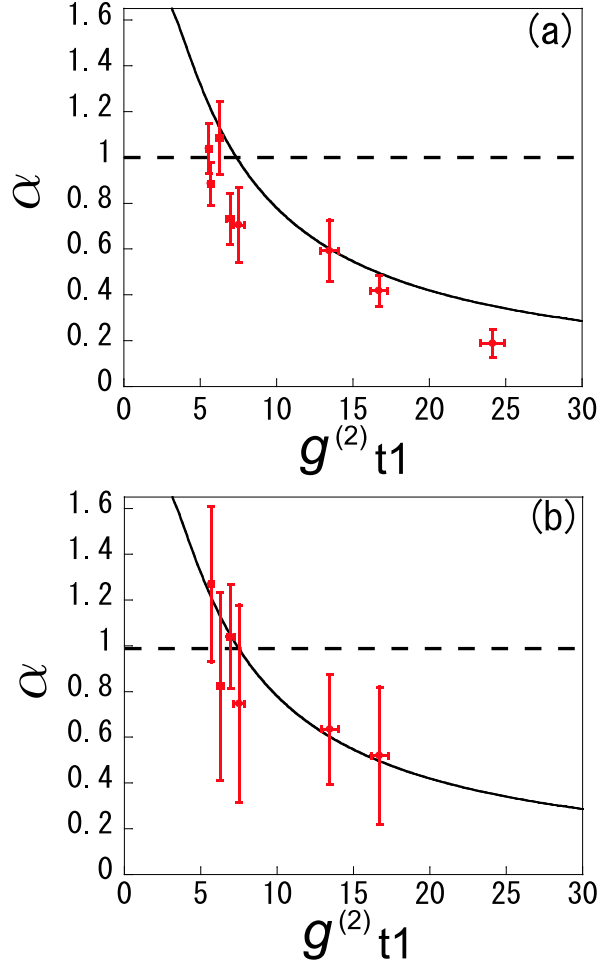}
  \caption{\label{fig:anti}  Obtained anticorrelation parameter of (a) the input frequency-filtered PDC photons and (b) the retrieved component as a function of the intensity correlation function of the input. The solid line is a simple theoretical curve. 
}
 \end{center}
\end{figure}

Fig. \ref{fig:anti} shows the anticorrelation parameter $\alpha$ for the (a) incident and (b) retrieved states 
as a function of $g^{(2)}_{t1}$ of the incident states, 
which is the cross intensity correlation function between SPCM-t and SPCM-1. 
The error bars indicate the standard deviations and 
the solid line shows a theoretical curve based on the degenerate parametric process. 
As the value of $g^{(2)}_{t1}$ increased, the both anticorrelation parameters for incident and retrieved states decreased. 
This property is essential for the PDC light to be the conditional single photon state. 
The triggered incident light showed a single photon property with a minimum value of $\alpha=0.19\pm0.06$. 
While the measurement error increased due to a decrease of the photon flux caused by an imperfect storage process, the triggered retrieved light showed nonclassical feature associated with a single photon state, where a minimum value of $\alpha=0.52\pm0.30$ was obtained. 
The probability that the $\alpha \ge 1$ is 0.08 (the probability for the point at $g^{(2)}_{t1}=13.5$ is 0.05). 
While the anticorrelation parameter for the retrieved light does not show enough small value to be considered as the single photon, the conditional retrieved state is in the nonclassical region. 
The incident light included noncorrelated components due to the Lorentzian shape of the filtered spectrum. The retrieved light also included the offset component. Nevertheless, as is shown in fig. \ref{fig:anti}, the anticorrelation parameters for both incident and retrieved light were in agreement with the simple theoretical curve without a fitting parameter. This is because the anticorrelation parameter was plotted as a function of $g^{(2)}_{t1}$, which is sensitive to the strength of the photon pair correlation in the light.

\section{Conclusion}
In conclusion, we have demonstrated storage and retrieval of PDC photons with EIT, where the bandwidth of the light was reduced from THz to MHz order. 
Storage and retrieval of the flux of the PDC photons were clearly observed due to the frequency filtering. 
We introduced an inequality which is only satisfied for the classical light field and we verified the nonclassicality of the retrieved light from the violation of this inequality. This result suggests that the atomic ensemble storing the photonic information was in a nonclassical state. 
We also prepared the conditional single photons by using the PDC photons and performed the storage and retrieval of the single photons.
While the anticorrelation parameter $\alpha$ for the retrieved light did not show the value close to 0, $\alpha$ decreased as $g^{(2)}_{t1}$ increased and showed the value less than 1 (non-classicality). 
Since the PDC photons have been widely utilized as the resource of nonclassical state of light, the results obtained here should have the wide range of applications which are not limited to the quantum information processing. For example, manipulation of nonclassical lights will be possible by controlling the state of atoms storing the photonic information. A variety of nonclassical atomic states also could be generated, which might be utilized for testing bizarre nature of quantum physics or measuring the fundamental physical constants with ultra high precision.

\ack
We gratefully acknowledge M. Koashi, D. Akamatsu, K. Usami and T. Yonehara.
One of the authors (K. A.) was partially supported by the JSPS.
This work was supported by a Grant-in-Aid for Scientific Research (B) and 
the 21st Century COE Program at Tokyo Tech ``Nanometer-Scale Quantum Physics'' by MEXT.

\end{document}